\documentclass[prl,aps,twocolumn,showpacs]{revtex4}

\usepackage{graphicx}
\graphicspath{{fig/}}
\usepackage{amsmath}

\newcommand{\s}{\sigma}
\newcommand{\sbar}{\bar{\s}}
\newcommand{\ds}{\delta\sigma}
\newcommand{\e}{\varepsilon}
\newcommand{\ebar}{\bar{\e}}
\newcommand{\de}{\delta\varepsilon}
\newcommand{\la}{\lambda}
\newcommand{\g}{\gamma_\nu}
\newcommand{\gbar}{\bar{\gamma}_\nu}

\begin{document}

\title{Jahn-Teller Distortions and the Supershell Effect in Metal Nanowires}

\author{D.~F.~Urban$^1$, J.~B\"urki$^{1,2}$, C.-H.~Zhang$^2$,
C.~A.~Stafford$^2$, and Hermann~Grabert$^1$}

\affiliation{${}^1$Physikalisches Institut, Albert-Ludwigs-Universit\"at, D-79104 Freiburg, Germany \\
${}^2$Department of Physics, University of Arizona, Tucson, AZ 85721}

\date{\today}

\begin{abstract}
A stability analysis of metal nanowires shows that a
Jahn-Teller deformation breaking cylindrical symmetry can be
energetically favorable, leading to stable nanowires with
elliptic cross sections. The sequence of stable cylindrical and
elliptical nanowires allows for a consistent interpretation of
experimental conductance histograms for alkali metals, including both
the electronic shell and supershell structures.  It is predicted
that for gold, elliptical nanowires are even more likely to form
since their eccentricity is smaller than for alkali metals.
The existence of certain metastable ``superdeformed'' nanowires is also predicted.
\end{abstract}


\maketitle \vskip2pc

%
%

The effects of shell filling on the abundance spectra of metal
clusters have been known for years and explain the existence of
clusters with ``magic numbers,'' corresponding to full electronic
shells, which are observed more frequently than others
\cite{Heer96}. More recently, electronic shell and supershell
structures in alkali metal nanowires have been reported by the
Leiden group \cite{Yanson99,Yanson00,Yanson01}. The conductance
$G$ was recorded during the breaking of nanowires in a
mechanically controllable break junction (MCBJ), and histograms were
built out of thousands of measurements. When the temperature is
large enough to allow the wire to explore the phase space of
possible shapes, the shell structure shows up as peaks in the
histogram of $\sqrt{G}$ that are equally spaced, while the
supershell structure manifests itself as a modulation of the
amplitude of the peaks (See Agra{\"\i}t \emph{et al.}
\cite{Agrait03} for a recent review).

The electron-shell structure in alkali nanowires can be understood qualitatively
using the free-electron model of a cylinder \cite{Yanson99}.
However, more detailed stability analyses of axisymmetric wires
\cite{Kassubek01,Zhang03,Urban03} have
revealed a sequence of stable ``magic'' radii, with characteristic gaps
that are not fully consistent
with the nearly perfect periodicity of the experimentally
observed peak positions.  The deviations can be
accounted for neither by the inclusion of disorder
\cite{Buerki01}, nor by the use of more elaborate
self-consistent jellium models \cite{Puska01}. Since gaps
in the sequence of cylindrical nanowires arise from a degeneracy
of conductance channels, it is natural to assume that a
Jahn-Teller deformation breaking the symmetry can lead to more
stable deformed configurations.

While the experimental manifestations of electron-shell structure
are similar in metal clusters and nanowires, the Jahn-Teller
effect plays out quite differently in these two systems. Metal
clusters and nanowires differ fundamentally in that surface
effects tend to stabilize clusters, while they lead to a Rayleigh
instability in nanowires \cite{Kassubek01}. As a result,
Jahn-Teller deformations of clusters are very common and typically
rather small \cite{bulgac93,Schmidt99} while, as we show here,
they only occur for a minority of nanowires and can be rather
large. Nevertheless, there are several stable nanowires with
elliptic cross-sections (referred to as \emph{elliptical
nanowires} henceforth), leading to a sequence of stable
cylindrical and elliptical nanowires that allows for a satisfying
interpretation of the experiments on electronic shell and
supershell structure \cite{Yanson99,Yanson00,Yanson01}. In
addition, our theory predicts the existence of short-lived
superdeformed wires.

%
%

We use a nanoscale free-electron model, treating the
electrons as a non-interacting Fermi gas confined within the wire
by hard-wall boundary conditions \cite{Stafford97a}. This
continuum model is especially suitable for alkali metals, but can
also be applied to other monovalent metals. Macroscopic arguments
\cite{Zhang03} suggest that the surface tension $\sigma_s$ should
drive a crossover from a crystalline solid to a plastic or fluid
below a critical radius $R_{\rm min}=\sigma_s/\sigma_Y$, where
$\sigma_Y$ is the yield strength.  Typically, $R_{\rm min} \sim
10\,\mbox{nm}$ in metals \cite{Zhang03}. This implies that
electronic effects should dominate over atomistic effects for
sufficiently small radii. Indeed, a crossover from atomic-shell
to electron-shell effects with decreasing radius is observed in
both metal clusters \cite{Martin96} and nanowires
\cite{Yanson01,Diaz03,Medina03,Mares04}, justifying {\it a
posteriori} the use of the nanoscale free-electron model in the later regime.

Guided by evidence of the existence of non-spherical
clusters \cite{bulgac93,Schmidt99}, we consider straight elliptical nanowires
\cite{footnote1} aligned along the $z$-axis.
The wire cross-section is characterized
by the ellipse's two major semi-axes $a$ and $b$, or
equivalently, by the area parameter
$\s^2=a\cdot b$ and the aspect ratio $\e=a/b$.
When a small perturbation, written as a Fourier series, is added to a
wire of length $L$, the geometry is characterized by the two
functions
\begin{eqnarray}
\label{eq.perturbation}
     \s(z) &=& \sbar + \la\sum_q\s_q e^{i q z}\equiv \sbar + \la\,\ds(z),
\nonumber\\     \e(z) &=& \ebar + \la\sum_q\e_q e^{i q z}\equiv \ebar +
\la\,\de(z),\end{eqnarray}
where the dimensionless parameter $\lambda$ sets the size of the
perturbation. The perturbation wave vector $q$ must be an integer
multiple of $2\pi/L$ and, since $\s(z)$ and $\e(z)$ are real, we have
$\s_{-q}=\s^*_q$ and $\e_{-q}=\e^*_q$.
The deformation must fulfill another constraint which comes from
the fact that, depending on material parameters, the deformed
wire tries to find a compromise between a volume conserving
deformation and one ensuring electroneutrality \cite{Stafford99}.
Here we use the general constraint
\begin{equation}
\label{eq.constraint}
    {\cal N} \equiv k_F^3{\cal V} - \eta (3\pi k_F^2{\cal A}/8 ) = \rm{const},
\end{equation}
where ${\cal V}$ is the volume of the wire, ${\cal A}$ its
surface area, and $k_F=2\pi/\lambda_F$ the Fermi wavevector. The
parameter $\eta$ can be adjusted so as to fix the value of the
effective surface tension to the material-specific value. In
particular, $\eta=0$ corresponds to a constant-volume constraint,
and $\eta=1$ is the constraint of constant (zero temperature) Weyl charge.

A nanowire is an open system, with electrons being injected and
absorbed by the leads, so that we need to calculate its grand
canonical potential $\Omega$ in order to determine the energetic
cost of a deformation.
A fully quantum mechanical stability analysis of cylindrical
nanowires \cite{Urban03} shows that, if the wires are short
enough, and/or the temperature is not too low, their stability is
essentially determined by their response to long wavelength
perturbations.
For these we can use the adiabatic approximation and decouple the
transverse and longitudinal motions yielding a series
of effectively one-dimensional problems, which can be solved with
the WKB approximation, in which $\Omega$ is given by
\cite{Stafford97a}
\begin{eqnarray}
\label{eq.omega}
    \Omega[T,L;\s,\e] &\!\!=&\!\!
    \int_0^\infty\!\!\!\!\text{d}E\left(-\frac{\partial f}{\partial
E}\right)        \;\Xi[E,L;\s,\e],
    \\
\label{eq.xi}
    \Xi[E,L;\s,\e]&\!\!=&\!\!\!
    -\frac{8E_F}{3\lambda_F}\!\!\int_0^L\!\!\!\text{d}z
\sum_\nu\left(\frac{E\!-\!E_\nu(\s,\e)}{E_F}\right)^{\!3/2}\!\!\!\!\!\!\!.\quad
\end{eqnarray}
%
%
Here $f=\left(1+\exp[(E-E_F)/k_BT]\right)^{-1}$ is the Fermi
function at temperature $T$, $E_F$ is the Fermi energy, and the
$E_\nu$'s are the transverse eigenenergies. The sum in
Eq.~(\ref{eq.xi}) runs over all open channels $\nu$, for which
$E_\nu(\s,\e)<E$. A factor of 2 accounts for spin degeneracy.
Note that at zero temperature Eq.~(\ref{eq.omega}) reduces to
$\Omega[T\!\!=\!0,L]=\Xi[E_F,L]$.

\begin{figure}[t]
    \begin{center}
    \includegraphics[width=0.9\columnwidth,draft=false]{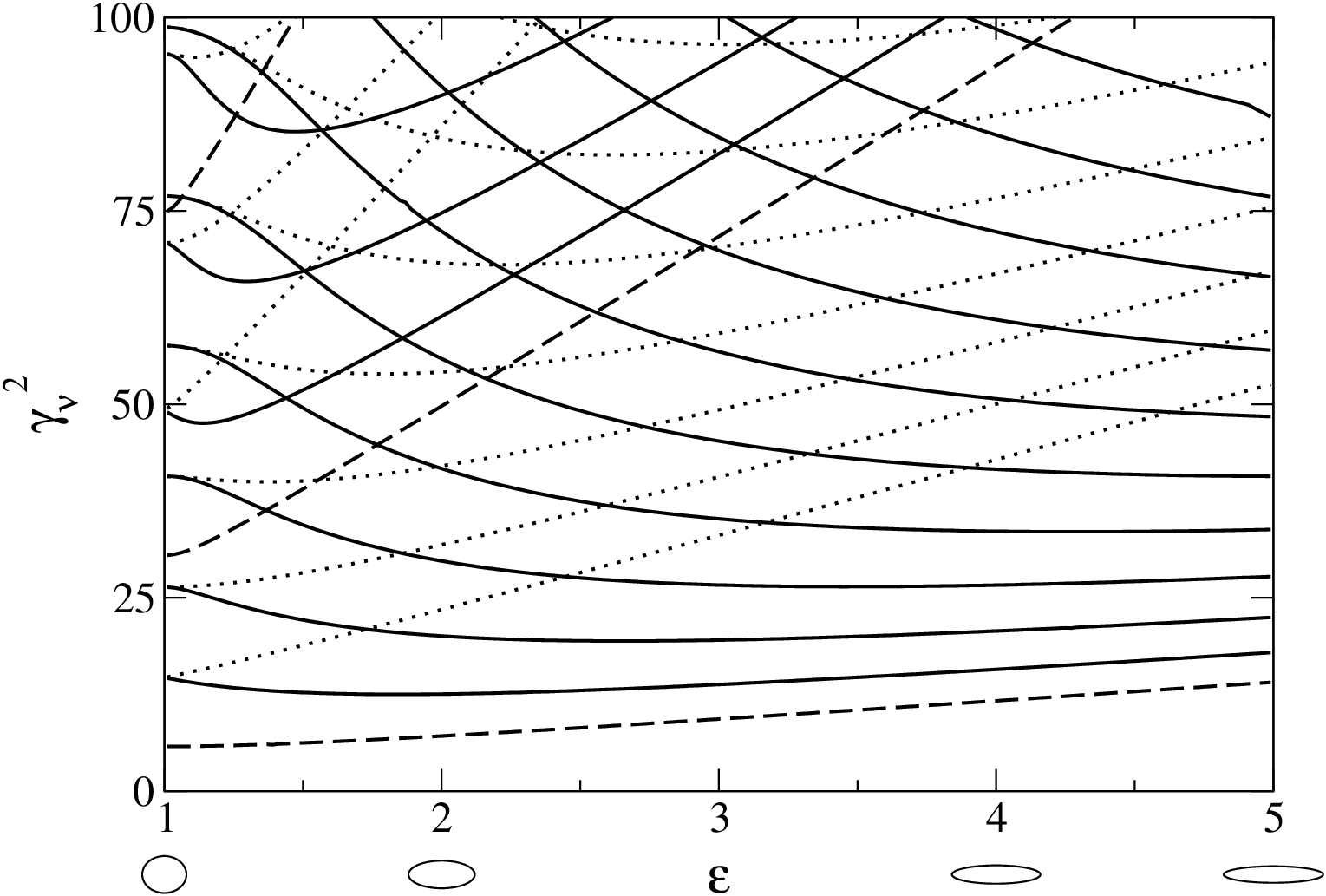}
    \end{center}
    \vspace{-0.5cm}
    \caption[]{\label{fig.eigenenergies} Eigenenergies of a 2d electron
            gas confined to an elliptical shape as a function of the aspect ratio $\e$.             Solid and dotted lines indicate even and odd states respectively,
        while states non-degenerate at $\e\!=\!1$ are shown with a dashed line.
        }
\end{figure}

The transverse eigenenergies can be written as
%
$E_\nu(\s,\e) = E_F\frac{\g(\e)^2}{(k_F\s)^2},$ where the $\g$'s
are the zeros of the modified Mathieu functions \cite{Broek01},
and can be computed numerically. In the limit $\e\rightarrow 1$,
we recover the result for a cylindrical wire, where most
eigenenergies are degenerate. As shown in
Fig.~\ref{fig.eigenenergies}, this degeneracy is lifted for
$\e\neq 1$ as a result of the decrease in symmetry, so that an elliptic nanowire
can sometimes be more stable than the axisymmetric one, leading
to a Jahn-Teller deformation.

The energetic cost of a small deformation of a straight elliptical
nanowire can be calculated by expanding Eq.~(\ref{eq.omega})
as a series in the parameter $\la$,
\begin{equation}
\label{eq.omega.expand}
    \Omega=\Omega^{(0)}+\la\Omega^{(1)}+\la^2\Omega^{(2)}+{\cal O}(\la^3).
\end{equation}
A nanowire $(\sbar,\ebar)$ is energetically stable at temperature $T$ if
$\Omega^{(1)}(\sbar,\ebar)=0$ and $\Omega^{(2)}(\sbar,\ebar)>0$ for
every possible deformation $(\de,\ds)$ satisfying $\partial{\cal{N}} = 0$.

%
%

The expansion of Eq.~(\ref{eq.xi}) yields
\begin{eqnarray}
\label{eq.omega0}
    \frac{\la_F}{L}\;\Xi^{(0)} &\!\!\!=&\!\! -\frac{8}{3\sqrt{E_F}}
    \sum_\nu\;
    [E-\bar{E}_\nu]^{3/2},
    \\
\label{eq.omega1}
    \frac{\la_F}{L}\;\Xi^{(1)} &\!\!\!=&\!\! \frac{8}{\sqrt{E_F}}
\sum_\nu    \bar{E}_\nu \sqrt{E-\bar{E}_\nu}
    \left(\frac{\gbar'}{\gbar}\,\e_0-\frac{\s_0}{\sbar}\right),
  \\
\label{eq.omega2}
  \frac{\la_F}{L}\;\Xi^{(2)}&\!\!\!=&\!\!\frac{4}{\sqrt{E_F}} \sum_q
    \binom{\s_q/\sbar}{\e_q}^{\!\!\dagger} \!\!
    \begin{pmatrix} A_{\s\s} & \!A_{\s\e} \\ A_{\s\e} &
\!A_{\e\e}\end{pmatrix}\!\!    \binom{\s_q/\sbar}{\e_q},\;\;\;
\end{eqnarray}
where $\bar{E}_\nu=E_\nu(\sbar,\ebar)$, and $\gbar=\g(\ebar)$. The
prime indicates differentiation with respect to $\e$. The
elements of the matrix $A$ in Eq. (\ref{eq.omega2}) are given by
\begin{eqnarray}
\label{eq.stabcoef}
  A_{\s\s}&\!\!=&\!\!\sum_\nu \bar{E}_\nu\left(
3\sqrt{E\!-\!\bar{E}_\nu}-\frac{\bar{E}_\nu}{\sqrt{E\!-\!\bar{E}_\nu}}
\right), \nonumber \\%
  A_{\s\e}&\!\!=&\!\!\sum_\nu
  \bar{E}_\nu\!\left[\!\!
\left\{\!\left[\!\frac{\gbar'}{\gbar}\!\right]^2\!\!\!+\frac{\gbar''}{\gbar}\!\right\}\!    \sqrt{\!E\!-\!\bar{E}_\nu\!}
-\left[\!\frac{\gbar'}{\gbar}\!\right]^2
\!\!\!\!\frac{\bar{E}_\nu}{\sqrt{\!E\!-\!\bar{E}_\nu}\!}
\right]\!,\nonumber \\%
  A_{\e\e}&\!\!=&\!\!-\sum_\nu \bar{E}_\nu
\frac{\gbar'}{\gbar}\left(
2\sqrt{E\!-\!\bar{E}_\nu}\!-\!\frac{\bar{E}_\nu}{\sqrt{E\!-\!\bar{E}_\nu
}}  \right).%
\end{eqnarray}
\begin{figure*}
    \begin{minipage}[c]{11.6cm}\hspace*{-6mm}
      \includegraphics[width=11.6cm,clip=,draft=false]{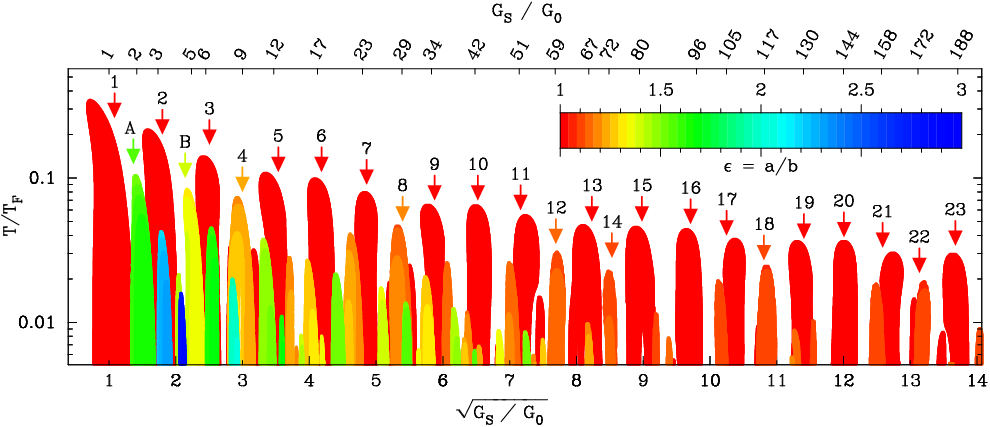}
    \end{minipage}\
    \begin{minipage}[c]{5.8cm}
      \caption[]{\label{fig.stabzone} (color)
      Energetically stable cylindrical and elliptical nanowires as a
      function of temperature (in units of the Fermi temperature $T_F$).
      The surface tension was adjusted to a value of $0.22\,\rm{N/m}$,
      corresponding to Na \cite{Tyson77}.
      The aspect ratio $\e$ is coded via the scale shown in the inset.
      The numbered arrows label the most stable configurations
      that enter our analysis of the shell structure. Arrows A and B label
      two highly deformed elliptical wires that are
      addressed specifically in the text.
      }
    \end{minipage}\vspace*{-3mm}
\end{figure*}
The constraint $\partial{\mathcal{N}}=0$ on allowed deformations
restricts the number of independent Fourier coefficients in Eq.
(\ref{eq.perturbation}). Hence we can express $\s_0$ in terms of
the other coefficients and expand it as a series in $\la$,
\begin{equation}
\label{eq.expands0}
\s_0 = \s_0^{(0)}\big(\ebar,\sbar,\e_0 \big) +
    \la\, \s_0^{(1)}\big(\ebar,\sbar,\{\e_q, \s_q\}\big) +
{\cal{O}}(\la^2),
\end{equation}
and eliminate $\s_0$ from Eqs.~(\ref{eq.omega1}) and (\ref{eq.omega2}),
modifying the expressions for $\Xi^{(1)}$ and the stability matrix $A$.
At zero temperature, the modified condition $\tilde{\Xi}^{(1)}=0$ determines which
wires are stationary states, while the positivity of the modified
matrix $\tilde{A}$ reveals the stability of the wire. The results
for finite temperature can be derived analogously by evaluating
Eq.~(\ref{eq.omega}) numerically.

Figure~\ref{fig.stabzone} shows the stable geometries as a function
of temperature and conductance, with the surface tension adjusted to the value $0.22$
N/m (i.e. $\eta=0.93$), corresponding to Na \cite{Tyson77}. The color represents the
value of the aspect ratio $\e$ and the $x$-axis is given by the
square root of the corrected Sharvin conductance
%
%
%
\mbox{$G_S=G_0(k_F^2\s^2\!/4\,-\,k_F{\cal{P}}\!/4\pi\,+\,1\!/6)$}.
Here $G_0=2e^2/h$ is the conductance quantum and ${\cal{P}}$ the
ellipse perimeter.

The possibility to observe a stable wire in conductance histograms
depends mostly on two conditions:
({\sl i}) The wire has to be formed often enough to be statistically
relevant;
({\sl ii}) It needs to have a long enough lifetime so as to be
recorded. Condition ({\sl i}) depends on the ability of the system to
probe new configurations, which is determined by the mobility of the
atoms and the density of stable geometries in configuration space.
Regarding point ({\sl ii}), a calculation of the lifetime of a nanowire
is beyond the scope of this paper, but one can
expect that wires that are linearly stable up to larger
temperatures $T_{max}$ will have longer lifetimes at low temperatures \cite{lifetime}.

\begin{figure}[b]
  \begin{center}
       \includegraphics[width=0.82\columnwidth,draft=false,clip=]{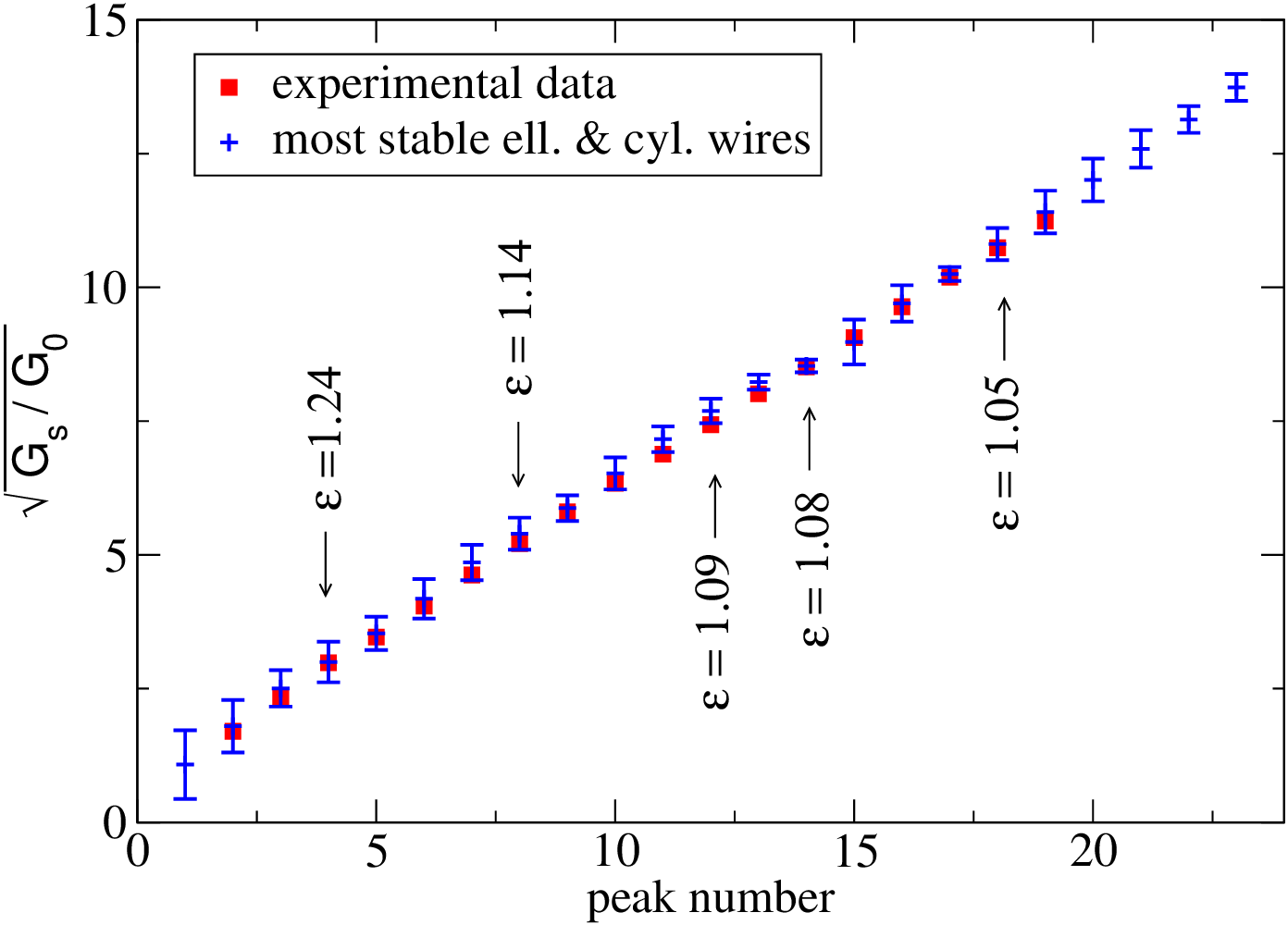}
  \end{center}
  \vspace{-0.5cm}
  \caption[]{\label{fig.shell}
    (color online) Comparison of the experimental shell structure
    for Na (taken from Ref.~\cite{Yanson99}) with our theoretical
    predictions of the most stable Na nanowires. Elliptical wires are
    labeled with the corresponding aspect ratio $\e$.}
\end{figure}

Based on these considerations, we extract the most stable
configurations from Fig.~\ref{fig.stabzone}, defined as the
geometries that persist up to the highest temperature compared to
their neighboring configurations. Those wires are marked with
numbered arrows in Fig.~\ref{fig.stabzone}. For each stability
peak, we extract its mean Sharvin conductance and its width, and plot
them as a function of the peak number in Fig.~\ref{fig.shell},
together with the experimental data from Ref.~\cite{Yanson99}.
Note that the striking fit is only possible when including elliptical
nanowires, for which the corresponding aspect ratios $\e$ are also
shown in Fig.~\ref{fig.shell}.

The heights of the dominant stability peaks in Fig.\ \ref{fig.stabzone}
exhibit a periodic modulation, with minima occuring at
peaks 4, 8, 12, 18, etc.  The positions of these minima are in
perfect agreement with the observed supershell structure in
conductance histograms of alkali metal nanowires \cite{Yanson00}.
Interestingly, the nodes of the supershell
structure, where the shell effect for a cylinder is suppressed, are precisely
where the most stable elliptical nanowires are predicted to occur.
Thus the Jahn-Teller distortions and the supershell effect are
inextricably linked.

While both the shell \cite{Yanson99} and supershell
\cite{Yanson00} effects are
accurately described by our stability analysis, our thermodynamic model does not
directly address the complex dynamical process by which various contact
structures form in MCBJ experiments, and therefore cannot describe every
detail of the conductance histograms, such as their highly non-trivial temperature dependence.

Our stability analysis also reveals two highly deformed
elliptical nanowires with conductance values of $2\;G_0$ and
$5\;G_0$, marked in Fig.\ \ref{fig.stabzone} by arrows A and B,
respectively. They are expected to appear more rarely due to
their reduced stability relative to the neighboring peaks, 
and their large aspect ratio $\e$ (see Table\ \ref{tab.wires})
that renders them rather isolated in configuration space
\cite{footnote2}. Conductance histograms of the alkali metals
often do show a conductance peak at $5\;G_0$ and a shallow
shoulder at $2\;G_0$ \cite{Yanson99,Yanson01}. Nevertheless,
these peaks were {\em not included} in the experimental analysis
of shell structure in Refs.\ \onlinecite{Yanson99} and
\onlinecite{Yanson01}. Note that the definition of an
experimental ``peak'' depends on the selection criteria and
smoothing of the data.

Of the three alkali
metals, Li, Na and K, Potassium has the lowest and Lithium the
highest melting temperature. This suggests that at a given
temperature the mobility of the atoms is highest in K and lowest
in Li. This is reflected in the fact that the evidence of the
highly deformed nanowires A and B is clear for K, can be seen for Na, but
is barely visible for Li \cite{Yanson01}.

In addition to wire A, Fig.\ \ref{fig.stabzone} shows a
number of ``superdeformed'' nanowires with $\e > 1.5$.  The most
stable are at $(G,\,\e)=$ (2, 1.65), (3, 2.25), (4, 2.8), (6,
1.65), (8, 2.0), (12, 1.65), (19, 1.55), (29, 1.55) and (41, 1.6).
An aspect ratio near 2 is favorable for the shell effect
\cite{SuperD}, as evidenced by the large gaps in the energy
spectrum (Fig.\ \ref{fig.eigenenergies}), but the large
cost in surface energy renders these wires less stable than the
magic cylindrical wires and the ``normal-deformed'' wires listed
in Table\ \ref{tab.wires}.

\begin{table}[t]
\begin{tabular}{|c|c|c|c|c|c|}
    \colrule
      & $G_S/G_0$ & $\;\e^{\rm{(Na)}}\;$ &
$\;k_BT_{\rm{max}}^{\rm{(Na)}}\;$ [eV]      & $\;\e^{\rm{(Au)}}\;$ &
$\;k_BT_{\rm{max}}^{\rm{(Au)}}\;$ [eV] \\    \colrule
A  &  2.3   &   1.65  &  0.360  &   1.50  &  0.376 \\
B  &  5.3   &   1.32  &  0.282  &   1.27  &  0.326 \\
4  &  9.0   &   1.24  &  0.243  &   1.17  &  0.288 \\
8  &  29.1  &   1.14  &  0.152  &   1.12  &  0.194 \\
12 &  59.2  &   1.09  &  0.100  &   1.08  &  0.122 \\
14 &  72.8  &   1.08  &  0.076  &   1.07  &  0.083 \\
18 &  116.2 &   1.05  &  0.081  &   1.05  &  0.105 \\
    \colrule
\end{tabular}
\caption{Most stable elliptical nanowires (with $\e>1$) for Na
and Au. Listed are the labels from Fig.\ \ref{fig.stabzone}, the
mean Sharvin conductance $G_S$, the aspect ratio $\e$, and the
maximum temperature $T_{\rm{max}}$ up to which the wires remain
stable, which is related to the depth of the corresponding
energetic minimum. }
\label{tab.wires}
\end{table}
Conductance histograms have also been recorded for gold, both at
low \cite{Costa97} and room temperatures
\cite{Diaz03,Medina03,Mares04}, where evidence of electronic shell
effects has been reported.
Although the explanation of some features of Au wires, e.g.
surface reconstruction, requires explicit inclusion of
5d-orbitals, the jellium model is sufficient to explain
electron-shell effects.
Performing a stability analysis for gold requires adjusting the
deformation constraint (\ref{eq.constraint}) according to the
surface tension for Au, ($1.3\,\rm{N/m}$ \cite{Tyson77}, $\eta=0.61$), which is higher than for Na.
Therefore the results are somewhat different: Stable
configurations appear at the same conductance values, but the
temperature $T_{\rm{max}}$ up to which elliptical wires remain
stable is larger, and these wires are less deformed than for Na
(see Table \ref{tab.wires}). As a consequence, the probability to
observe elliptical nanowires in experiments is enhanced. Indeed,
several experimental histograms \cite{Costa97,Diaz03} for Au
show clear peaks at conductance values of $2\;G_0$ and $5\;G_0$.

In conclusion, we have presented a stability analysis of
elliptical metal nanowires, using a jellium model, and have shown
that Jahn-Teller-distorted wires can be stable. The derived
sequence of stable cylindrical and elliptical geometries explains
the experimentally observed shell and supershell structures for
alkali metals. Deformed wires can explain additional conductance
peaks observed in alkali metals and gold.

We are grateful to J. van Ruitenbeek for sharing his data on
Na conductance histograms.
This research has been supported by the DFG through SFB 276 and
the EU Network DIENOW.
CHZ and CAS acknowledge support from NSF grant DMR0312028.
\nocite{Buerki04}

\bibliography{Urban04_final}

\begin{thebibliography}{26}
\bibitem[{foo({\natexlab{a}})}]{footnote1}
\bibinfo{howpublished}{Higher-order multipole deformations are also possible,
  but cost significantly more surface energy. They are therefore only expected
  to occur at low conductance.}
\bibitem[{lif()}]{lifetime}
\bibinfo{howpublished}{Recent calculations for axisymmetric wires
  \cite{Buerki04} confirm this for neighboring peaks, but show that the
  relative lifetimes of wires with very different conductances cannot be
  directly inferred from $T_{max}$.}
\bibitem[{foo({\natexlab{b}})}]{footnote2}
\bibinfo{howpublished}{A nanowire produced by pulling apart an axisymmetric
  contact has a smaller probability to transform into a highly deformed
  configuration than into a neighboring cylindrical configuration.}
\end{thebibliography}

\end{document}